\documentclass[twocolumn,showpacs]{revtex4}
\usepackage{graphicx}
\usepackage{dcolumn}
\usepackage{bm}
\usepackage{amsmath}
\usepackage{amsfonts}
\usepackage{amssymb}

\providecommand{\U}[1]{\protect\rule{.1in}{.1in}}

\begin{document}

\title{Driving quantum walk spreading with the coin operator}
\author{A. Romanelli}
\altaffiliation{alejo@fing.edu.uy}
\affiliation{Instituto de F\'{\i}sica, Facultad de Ingenier\'{\i}a\\
Universidad de la Rep\'ublica\\
C.C. 30, C.P. 11000, Montevideo, Uruguay}
\date{\today }

\begin{abstract}
We generalize the discrete quantum walk on the line using a time dependent
unitary coin operator. We find an analytical relation between the long-time
behaviors of the standard deviation and the coin operator. Selecting the
coin time sequence allows to obtain a variety of predetermined asymptotic
wave-function spreadings: ballistic, sub-ballistic, diffusive, sub-diffusive
and localized.
\end{abstract}

\pacs{03.67.-a, 05.45.Mt}
\maketitle

\section{Introduction}

The quantum walk (QW) has been studied as a natural generalization of the
classical random walk in relation with quantum computation and quantum
information processing \cite{Kempe,Aharonov}. The possibility that future
quantum algorithms will be based on the QW has attracted the attention of
researchers from different fields. This topic is a matter of continuous
attention, as an example it has been recently suggested by Childs \cite%
{childs} that QW can be regarded as a universal quantum computation. Some
possible experimental implementations of the QW have been proposed by a
number of authors \cite{Dur,Travaglione,Sanders,Knight}, beyond them, the
fact remains that the development of experimental techniques to trap samples
of atoms using resonant exchanges of momentum and energy between atoms and
laser light \cite{cohen} may provide a realistic frame to implement quantum
computers \cite{Cirac}.

One of the most striking properties of the one-dimensional QW is its ability
to spread over the line linearly in time as characterized by the standard
deviation $\sigma (t)\sim t$, while its classical analog spreads out as the
square root of time $\sigma (t)\sim t^{1/2}$.

Several authors have studied the QW subjected to different types of coin
operators and/or sources of decoherence to analyze and verify the principles
of quantum theory as well as the passage from the quantum to the classical
world. In Ref. \cite{Tregenna1} it was shown that the phases in the coin
operator and the initial state of the coin can be used to control the
evolution of the QW. The presence of decoherence in the QW has been studied
as one possible route to classical behavior \cite{Brun1, Tregenna2,kendon}.
On the other hand the appearance of a small decoherence can be used to
enhance some properties of the QW in the development of quantum algorithms
\cite{Tregenna2}. The QW subjected to multiple independent coins was
analyzed in Ref. \cite{Brun2}, showing that this type of excitation can lead
to a diffusive spreading instead of the ballistic one. The QW with a
position dependent phase is studied in Refs. \cite{Luczak, auyuanet} where
dynamical localization was found. A coin with explicit time-dependence is
introduced for the first time in Ref. \cite{carlos1}, also finding dynamical
localization and quasi-periodic dynamics. In Ref. \cite{carlos2} a
non-linear dependence on the position-chirality probabilities is introduced
in the evolution, finding a variety of dynamical behaviors, including
ballistic motion and dynamical localization. The QW was also generalized
\cite{Ribeiro} introducing two coin operators arranged in quasi-periodic
sequences following a Fibonacci prescription, leading to a sub-ballistic
wave function spreading, as shown by the power-law tail of the standard
deviation [$\sigma (t)\sim t^{c}$ with $0.5<c<1$]. Simultaneously, we
studied the QW subjected to measurements \cite{alejo1} and decoherence \cite%
{alejo2} with a L\'{e}vy waiting-time distribution, finding that the system
has also a sub-ballistic spreading.

The appearance of this type of spreading shows the existence of a wealth of
quantum behaviors that are not restricted to be diffusive (decoherent) or
ballistic (coherent). These are quantum coherent intermediate situations
that could be useful to control quantum information. Also, to drive the
speed of the wave function spreading could be a tool for the development of
quantum algorithms. In Refs. \cite{alejo1,alejo2} some analytical results
were obtained about sub-ballistic behavior in a stochastic frame, but they
cannot be extended to deterministic cases such as in Ref. \cite{Ribeiro}.
Then it remains to clarify how some deterministic sequences of the coin
operator lead to a behavior that is neither diffusive nor ballistic. Here we
develop a simple QW model with a generalized coin, that allows an analytical
treatment, to clarify the connection between the time evolution of the coin
and the type of spreading of the system. We show how to select the sequences
of the coin operator to obtain a predetermined power-law distribution.

The paper is organized as follows. In the next section we develop the QW
model with a time depended coin, in the third section numerical results are
presented, in the last section we draw the conclusions, and in the appendix
A we show the analytical deduction for the moments.

\section{QW with time dependent coin}

\subsection{The standard QW}

The standard QW corresponds to a one-dimensional evolution of a quantum
system (the walker) in a direction which depends on an additional degree of
freedom, the chirality, with two possible states: \textquotedblleft
left\textquotedblright\ $|L\rangle $\ or \textquotedblleft
right\textquotedblright\ $|R\rangle $. The global Hilbert space of the
system is the tensor product $H_{s}\otimes H_{c}$ where $H_{s}$ is the
Hilbert space associated to the motion on the line and $H_{c}$ is the
chirality Hilbert space. Let us call $T_{-}$ ($T_{+}$) the operators in $%
H_{s}$ that move the walker one site to the left (right), and $|L\rangle
\langle L|$ and $|R\rangle \langle R|$ the chirality projector operators in $%
H_{c}$. We consider the unitary transformations
\begin{equation}
U(\theta )=\left\{ T_{-}\otimes |L\rangle \langle L|+T_{+}\otimes |R\rangle
\langle R|\right\} \circ \left\{ I\otimes K(\theta )\right\} ,  \label{Ugen}
\end{equation}%
where $K(\theta )=\sigma _{z}e^{-i\theta \sigma _{y}}$, $I$ is the identity
operator in $H_{s}$, and $\sigma _{y}$ and $\sigma _{z}$ are Pauli matrices
acting in $H_{c}$. The unitary operator $U(\theta )$ evolves the state in
one time step $\tau $ as $|\Psi (t+\tau )\rangle =U(\theta )|\Psi (t)\rangle
$. The wave vector can be expressed as the spinor
\begin{equation}
|\Psi (t)\rangle =\sum\limits_{k=-\infty }^{\infty }\left[
\begin{array}{c}
a_{k}(t) \\
b_{k}(t)%
\end{array}%
\right] |k\rangle ,  \label{spinor}
\end{equation}%
where the upper (lower) component is associated to the left (right)
chirality. The unitary evolution implied by Eq.(\ref{Ugen}) can be written
as the map
\begin{align}
a_{k}(t+\tau )& =a_{k+1}(t)\,\cos \theta \,+b_{k+1}(t)\,\sin \theta \,
\label{mapa0} \\
b_{k}(t+\tau )& =a_{k-1}(t)\,\sin \theta \,-b_{k-1}(t)\,\cos \theta .
\label{mapa}
\end{align}

\subsection{Generalized QW}

Here we are generalizing the QW to the case where different quantum coins
are applied. In particular we consider a deterministic angular time
dependence $\theta =\theta (t)$ for the coin operator.

In order to uncouple the chirality components in Eqs.(\ref{mapa0},\ref{mapa}%
) we consider two consecutive time steps and then the original map can be
rearranged to obtain
\begin{eqnarray}
&&a_{k}(t+\tau )\,\sin \theta _{-}-a_{k}(t-\tau )\,\sin \theta \,  \notag \\
&=&a_{k+1}(t)\,\cos \theta \,\sin \theta _{-}-a_{k-1}(t)\,\sin \theta \cos
\theta _{-},  \label{eq1}
\end{eqnarray}%
\begin{eqnarray}
&&b_{k}(t+\tau )\,\sin \theta _{-}-b_{k}(t-\tau )\,\sin \theta \,  \notag \\
&=&b_{k+1}(t)\,\cos \theta \,\sin \theta _{-}-b_{k-1}(t)\,\sin \theta \cos
\theta _{-},  \label{eq2}
\end{eqnarray}%
where $\theta _{-}=\theta (t-\tau )$. Both components of the chirality
satisfy the same equation, then in the following we shall work with $%
a_{k}(t) $ only. We are interested in a time scale much larger than $\tau $.
Then making a development around the time $t$ in Eq.(\ref{eq1}) and
retaining the terms up to the first order in $\tau$ we have
\begin{eqnarray}
&& 2\frac{da_{k}(t)}{dt}\tau \sin \theta -a_{k}(t)\frac{d\theta }{dt}\tau
\cos \theta \,  \notag \\
&\simeq&\left( a_{k+1}-a_{k-1}\right) \cos \theta \sin \theta  \notag \\
&&- \frac{d\theta }{dt} \tau\left( a_{k+1} \cos ^{2}\theta-a_{k-1} \sin
^{2}\theta \right) .  \label{cuatro}
\end{eqnarray}
In order to make analytical calculations we shall restrict to a continuous
coin function $\theta (t)$ that satisfies the condition $d\theta
(t)/dt\rightarrow 0$ for $t\rightarrow \infty $. Then Eq.(\ref{cuatro})
reduces to
\begin{equation}
2\frac{da_{k}}{dt}\simeq \frac{\cos \theta }{\tau }\text{ }\left(
a_{k+1}-a_{k-1}\right) .  \label{mapa2}
\end{equation}%
We define now the effective dimensionless time
\begin{equation}
t^{\ast }=\frac{1}{\tau }\int\limits_{t_{0}}^{t}\cos \theta dt,  \label{time}
\end{equation}%
where $t>t_{0}>>\tau$ and $t_{0}$ is chosen in such a way that starting from
it, Eq.(\ref{mapa2}) reproduces correctly the asymptotic behavior of the
original map Eqs.(\ref{mapa0},\ref{mapa}). Then $t_{0}$ is taken as the
initial time of the process. Changing to the variable $t^{\ast}$ Eq.(\ref%
{mapa2}) becomes
\begin{equation}
2\frac{da_{k}}{dt^{\ast }}\simeq a_{k+1}-a_{k-1}.  \label{mapa3}
\end{equation}%
Note that this is the recursion relation satisfied by the cylindrical Bessel
functions, thus the general solution of Eq.(\ref{mapa3}), introducing the
initial amplitudes $a_{l}^{0}=a_{l}(t_{0})=a_{l}(t^{\ast}=0)$, is
\begin{equation}
a_{k}(t^{\ast })\simeq \sum\limits_{l=-\infty }^{\infty }\left( -1\right)
^{k-l}a_{l}^{0}\text{ }\,J_{k-l}(t^{\ast }),  \label{solua}
\end{equation}%
where $J_{l}$ is the $l$th order cylindrical Bessel function. In a similar
way, the solution for the same equation for $b(t)$ is
\begin{equation}
b_{k}(t^{\ast })\simeq \sum\limits_{l=-\infty }^{\infty }\left( -1\right)
^{k-l}b_{l}^{0}\text{ }\,J_{k-l}(t^{\ast }),  \label{solub}
\end{equation}%
with $b_{l}^{0}=b_{l}(t_{0})$. The initial conditions of these equation are
not necessarily the same as those used in the discrete map Eqs.(\ref{mapa0}, %
\ref{mapa}), because the approximation of a difference by a derivative does
not hold for small times. In this context a long time implies many
applications of the discrete map. The results Eqs.(\ref{solua}, \ref{solub}%
), with appropriate initial conditions, provide a good asymptotic
description for the dynamics of the discrete Eqs.(\ref{mapa0}, \ref{mapa}),
as we shall see below.

Additionally Eqs.(\ref{solua}, \ref{solub}) show that the long-time
propagation speed of the probability amplitudes is given by $%
v(t)=dt^*/dt=\cos \theta /\tau $.

The probability distribution for the walker's position at time $t^{\ast }$
is given by
\begin{equation}
P_{k}(t^{\ast })\simeq |a_{k}(t^{\ast })|^{2}+|b_{k}(t^{\ast })|^{2}\text{,}
\label{prob}
\end{equation}%
that can be expressed as
\begin{equation}
P_{k}(t^{\ast })\simeq \sum_{j,l=\infty }^{\infty }\left( -1\right)
^{-(j+l)}\left( a_{l}^{0}a_{j}^{0\ast }+b_{l}^{0}b_{j}^{0\ast }\right)
J_{k-l}(t^{\ast })J_{k-j}(t^{\ast })\text{.}  \label{prob1}
\end{equation}%
We calculate the first and second moments of this distribution using the
properties of the Bessel functions (see appendix A), obtaining:
\begin{equation}
m_{1}(t^{\ast })\simeq -t^{\ast }\sum_{j=-\infty }^{\infty }\Re \left[
a_{j}^{0}a_{j-1}^{0\ast }+b_{j}^{0}b_{j-1}^{0\ast }\right] +m_{1}(0)\text{,}
\label{mome0}
\end{equation}%
\begin{eqnarray}
m_{2}(t^{\ast }) &\simeq &\frac{(t^{\ast })^{2}}{2}\left( 1+\sum_{j=-\infty
}^{\infty }\Re \left[ a_{j}^{0}a_{j-2}^{0\ast }+b_{j}^{0}b_{j-2}^{0\ast }%
\right] \right)  \notag \\
&&-t^{\ast }\sum_{j=-\infty }^{\infty }\left( 2j-1\right) \Re \left[
a_{j}^{0}a_{j-1}^{0\ast }+b_{j}^{0}b_{j-1}^{0\ast }\right]  \notag \\
&&+m_{2}(0)\text{,}  \label{mome}
\end{eqnarray}%
where $\Re \left[ x\right] $ is the real part of $x$ and $m_{1}(0)$ and $%
m_{2}(0)$ are the moments at time $t^{\ast }=0$ ($t=t_{0}$). Note that, if
we take symmetrical initial conditions, the first moment vanishes due to the
the symmetrical dependence on the initial condition \cite{konno}. The
asymptotic behavior of the standard deviation $\sigma =\sqrt{m_{2}-m_{1}^{2}}
$ is
\begin{equation}
{\sigma }(t^{\ast })\simeq \sqrt{A\left( t^{\ast }\right) ^{2}+Bt^{\ast }+C},
\label{sigma0}
\end{equation}%
where $A$ $,$ $B$ and $C$ depend on the initial conditions. From Eq.(\ref%
{sigma0}) we conclude that if the effective time $t^{\ast }$ remains bounded
for all $t$, then the standard deviation remains bounded too. In this case
the distribution does not spread and is therefore localized. On the other
hand, when $t^{\ast }$ is unbounded $\sigma $ increases with $t^{\ast }$ and
the system diffuses. Then, in the asymptotic regime, with $t^{\ast }$
unbounded, we substituted the Eq. (\ref{sigma0}) by the following equation
\begin{equation}
{\sigma }(n)\simeq \sqrt{A}t^{\ast }.  \label{sigma1}
\end{equation}

\subsection{Choosing the coin operator}

Among the alternatives for the time dependence of $\theta (t)$ we study the
case
\begin{equation}
\cos \theta (t)=\frac{1}{\sqrt{2}}\left( \frac{\tau }{t+\tau }\right)
^{\alpha },  \label{cos}
\end{equation}%
where $\alpha \geq 0$. Substituting Eq.(\ref{cos}) into Eq.( \ref{time}) and
the resulting expression into Eq.(\ref{sigma1}) for $\alpha \leq 1$ \ and
using the Eq.(\ref{sigma0}) for $\alpha >1$, one obtains
\begin{equation}
{\sigma }(n)\simeq \left\{
\begin{array}{c}
\frac{1}{\left( 1-\alpha \right) }\sqrt{\frac{A}{2}}\left( n^{1-\alpha }-{%
n_{0}}^{1-\alpha }\right) \text{, if }\alpha <1\text{,} \\
\\
\sqrt{\frac{A}{2}}\ln \left( \frac{n}{n_{0}}\right) \text{, if }\alpha =1%
\text{,} \\
\\
\sqrt{C}\text{, if }\alpha >1,%
\end{array}%
\right.  \label{sigmaf}
\end{equation}%
where we have introduced the discrete time dimensionless time $n$ through
the definition $t=(n-1)\tau $ and $t_{0}=(n_{0}-1)\tau $. The standard
deviation, in the asymptotic limit $n>>n_{0}>>1$, is given by
\begin{equation}
{\sigma }(n)\rightarrow \left\{
\begin{array}{c}
n^{1-\alpha }\text{, if }\alpha <1\text{,} \\
\\
\ln n\text{, if }\alpha =1\text{,} \\
\\
1\text{, if }\alpha >1\text{,}%
\end{array}%
\right.  \label{sigma}
\end{equation}%
where we have omitted multiplicative constants.
\begin{figure}[th]
\begin{center}
\includegraphics[scale=0.38]{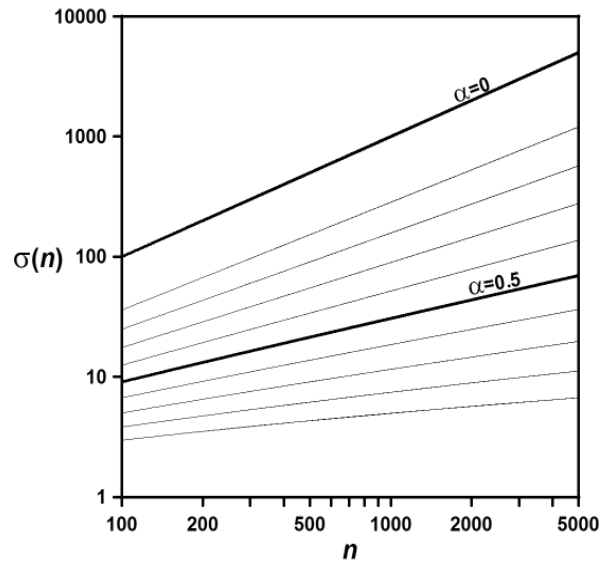}
\end{center}
\caption{The standard deviation ${\protect\sigma }(n)$ as a function of the
dimensionless time $n$ in log-log scales. The values of $\protect\alpha$
vary from $0$ (top) to $0.9$ (bottom) in steps of $0.1$. The curves for the
ballistic ($\protect\alpha=0$) and the diffusive ($\protect\alpha=0.5$)
cases are shown in thick line to help visualization}
\label{f1}
\end{figure}
\begin{figure}[th]
\begin{center}
\includegraphics[scale=0.38]{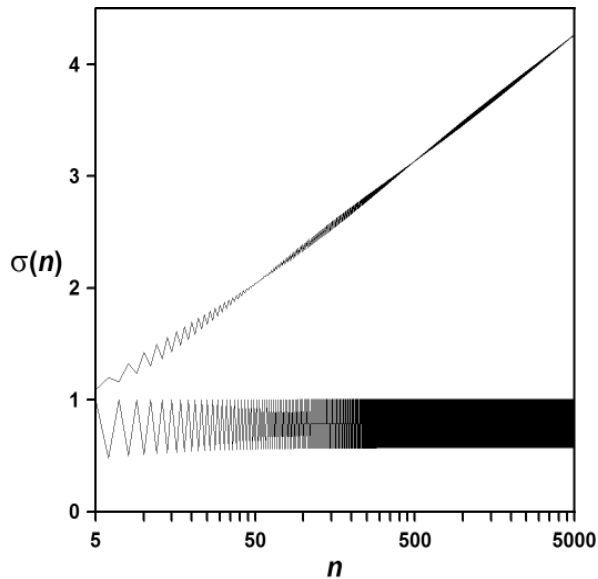}
\end{center}
\caption{The standard deviation $\protect\sigma(n)$ (in linear scale) as a
function of the dimensionless time $n$ (in logarithmic scale). The upper
curve for $\protect\alpha=1$ satisfies the law ${\protect\sigma }(n)\sim \ln
n$ for large $n$. The lower curve for $\protect\alpha=2$ oscillates around a
constant value.}
\label{f2}
\end{figure}
\begin{figure}[th]
\begin{center}
\includegraphics[scale=0.38]{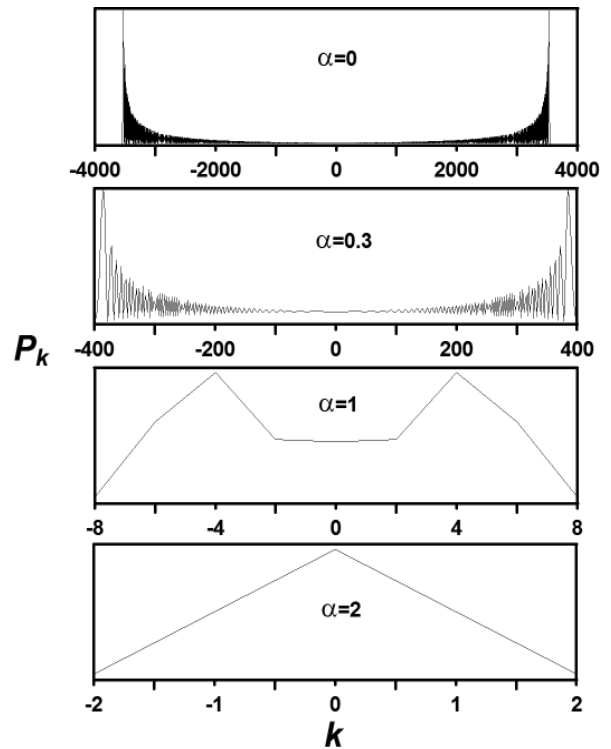}
\end{center}
\caption{The probability distribution $P_k$ as a function of the
dimensionless spatial position $k$ at time $t=5000\,\protect\tau$. The
values of the parameter are, from top to bottom, $\protect\alpha=0,\, 0.3,\,
1,$ and $2$. The scales for $k$ are different in each case}
\label{f3}
\end{figure}
This result predicts five different types of asymptotic behaviors:

a) ballistic for $\alpha =0$,

b) sub-ballistic for $0<\alpha <0.5$,

c) diffusive $\alpha =0,5$

d) sub-diffusive for $0.5<\alpha\leq 1$,

e) localized for $\alpha >1$. \newline
The result for $\alpha =0$ is to be expected because then the system reduces
to the usual QW with a Hadamard coin. In the other cases the system shows a
variety of behaviors where the standard deviation has a slower growth. The
most extreme of them arises for $\alpha >1$. In this case the system does
not spread when $t\rightarrow \infty $ and it maintains a localized position.

It should be noted that when the derivative of the coin function satisfies $%
d\theta (t)/dt\lesssim \tau$ for $t\rightarrow \infty$, Eq.(\ref{mapa2})
still correctly reproduces the asymptotic behavior of the original map.
Indeed, the approximation made to obtain Eq.(\ref{mapa2}) remains valid. In
particular, if we take the linear time dependence $\theta=2\pi \gamma t$,
with $\gamma$ irrational and $\gamma\lesssim \tau/(2\pi)$, the present model
gives similar results as those found in Ref.\cite{carlos1}.

\section{Numerical calculation}

In order to verify the approximations made in our analytical treatment we
shall compare the result of Eq.(\ref{sigma}) with the numerical evaluation
using Eqs.(\ref{mapa0}, \ref{mapa}). We take as initial conditions a walker
starting from the central position $|0\rangle $ with chirality $\frac{1}{%
\sqrt{2}}\left[ 1,i\right] ^{T}$; this choice produces a symmetrical
evolution as in the usual QW.

We then calculate the standard deviation $\sigma \left( n\right)$ using the
original map. The results for $\alpha$ from $0$ to $0.9$ are shown in Fig.~%
\ref{f1}, where for large $n$ ($t>1000\, \tau$) the curves have no
perceptible difference with $n^{1-\alpha}$.

In Fig.~\ref{f2} the upper curve, that corresponds to the standard deviation
for $\alpha=1$, shows some quick oscillations governed by an average
logarithmic evolution as was numerically checked for large times. The lower
curve, for $\alpha=2$, shows that the standard deviation oscillates around a
constant value. Then in both cases the average behavior agrees with the
theoretical predictions.

The profiles of the distribution $P_k$ for several values of $\alpha$ are
shown in Fig.~\ref{f3} at the same time $t=5000\,\tau$. For $\alpha=0$ the
known QW profile for the Hadamard coin is obtained. In this case the
function spreads with its greatest speed arriving up to the positions $%
\pm3500$. For larger values of $\alpha$ the spreading speed decreases, the
distribution shrinks and the two extreme peaks of the distribution come
closer. When $\alpha=0.5$ the standard deviation spreads out as $\sigma
(t)\sim t^{1/2}$, but the probability distribution $P_k$ is not Gaussian, it
has the same characteristics as shown in Fig.~\ref{f3} for $\alpha=0.3$.
This confirms that the evolution corresponds to a coherent unitary process
and not to the classical random walk. In the extreme case when $\alpha>1$
the distribution is restricted to small values around its initial value and
the two extreme peaks have melt in a very narrow peak. We conclude that in
all cases the asymptotic behaviors have an excellent agreement with the
theoretical predictions.

\section{Conclusion}

In the usual QW the standard deviation has a linear growth, however using a
biased quantum coin arranged in aperiodic sequence a sub-ballistic behavior
is obtained \cite{Ribeiro}. Here we generalize the discrete QW using a time
dependent unitary coin operator. We find an analytical expression for the
dependence of the standard deviation on the coin operator in the asymptotic
regime. This expression shows explicitly how to select the unitary evolution
to obtain a predetermined asymptotic behavior for the wave function
spreading: ballistic, sub-ballistic, diffusive, sub-diffusive or localized.
The parameter $\alpha $ of Eq.(\ref{cos}) determines the degree of
diffusivity of the system for large times. This feature opens interesting
possibilities for quantum information processing because it could be used
for controlling the spreading of the wave function.

Finally another point worth mentioning is that the asymptotic differential
Eq.(\ref{mapa3}) has the same form that the differential equation for the
amplitudes of the quantum kicked rotor (QKR) in resonance \cite{Izrailev}.
The similitude between the QW and the QKR in resonance, suggested in Refs.
\cite{alejo1,auyuanet,alejo6}, has been recently extended to the anomalous
diffusion behavior presented in this work \cite{NewQKR}. Therefore, as the
QKR has been realized experimentally \cite{Moore} and the quantum resonances
have already been observed \cite{Bharucha,Kanem}, our results for the QW
could be experimentally confirmed through an implementation of the QKR in
resonance.

We acknowledge the support from PEDECIBA, ANII and thank V. Micenmacher for
his comments and stimulating discussions.

\appendix
\section{}
 The first and second moments of the distribution $P_{k}$ are
\begin{eqnarray}
m_{1}(t) &=&\sum_{k=-\infty }^{\infty }kP_{k}(t),  \label{unop0} \\
m_{2}(t) &=&\sum_{k=-\infty }^{\infty }k^{2}P_{k}(t).  \label{dos0}
\end{eqnarray}%
Substituting Eq.(\ref{prob1}) into these equations we have%
\begin{eqnarray}
m_{1}(t) &=&\sum_{j,l=\infty }^{\infty }\left( -1\right) ^{-(j+l)}\left(
a_{l}^{0}a_{j}^{0\ast }+b_{l}^{0}b_{j}^{0\ast }\right) F_{jl},  \label{unop}
\\
m_{2}(t) &=&\sum_{j,l=\infty }^{\infty }\left( -1\right) ^{-(j+l)}\left(
a_{l}^{0}a_{j}^{0\ast }+b_{l}^{0}b_{j}^{0\ast }\right) {G}_{jl},
\label{dosp}
\end{eqnarray}%
where
\begin{eqnarray}
F_{jl} &=&\sum_{k=-\infty }^{\infty }k\text{ }J_{k-l}(t)J_{k-j}(t),
\label{s10} \\
G_{jl} &=&\sum_{k=-\infty }^{\infty }k^{2}\text{ }J_{k-l}(t)J_{k-j}(t).
\label{s20}
\end{eqnarray}%
With the change of indexes $\mu=k-l$ and $\nu =l-j$, these equations become
\begin{eqnarray}
{F}_{jl} &=&\sum_{\mu=-\infty }^{\infty }\mu\text{ }J_{\mu}(t)J_{\mu-\nu }(t)
\notag \\
&&+l\sum_{\mu=-\infty }^{\infty }J_{\mu}(t)J_{\mu-\nu }(t),  \label{s11} \\
{G}_{jl} &=&\sum_{\mu=-\infty }^{\infty }\mu^{2}\text{ }J_{\mu}(t)J_{\mu-\nu
}(t)  \notag \\
&&+2l\sum_{\mu=-\infty }^{\infty }\mu\text{ }J_{\mu}(t)J_{\mu-\nu }(t)
\notag \\
&&+l^{2}\sum_{\mu=-\infty }^{\infty }J_{\mu}(t)J_{\mu-\nu }(t).  \label{s22}
\end{eqnarray}%
In the above equations, three different type of sums are involved. They can
be expressed in terms of the Kronecker delta using the propertied of the
Bessel functions (Ref.\cite{Gradshteyn}, p. 992, Eq. \textbf{8.530})
\begin{eqnarray}
\sum_{\mu=-\infty }^{\infty }J_{\mu}(t)J_{\mu-\nu }(t) &=&\delta _{\nu 0},
\label{a} \\
\sum_{\mu=-\infty }^{\infty }\mu\text{ }J_{\mu}(t)J_{\mu-\nu }(t) &=&\frac{t%
}{2}\left( \delta _{\nu -1}+\delta _{\nu 1}\right) ,  \label{b} \\
\sum_{\mu=-\infty }^{\infty }\mu^{2}J_{\mu}(t)J_{\mu-\nu }(t) &=&\left(
\frac{t}{2}\right) ^{2}\left( \delta _{\nu -2}+2\delta _{\nu 0}+\delta _{\nu
2}\right)  \notag \\
&&+\frac{t}{2}\left( \delta _{\nu -1}-\delta _{\nu 1}\right) .  \label{d}
\end{eqnarray}%
Substituting the above equations into Eqs.(\ref{s11}, \ref{s22}) and
returning to the original indexes
\begin{eqnarray}
{F}_{jl} &=&l\text{ }\delta _{l j}+\frac{t}{2}\left( \delta _{l j-1}+\delta
_{l j+1}\right) ,  \label{s13} \\
{G}_{jl} &=&\left( \frac{t}{2}\right) ^{2}\delta _{l j-2}+\frac{t}{2}\left(
2l+1\right) \delta _{l j-1}  \notag \\
&&+\left[ 2\left( \frac{t}{2}\right) ^{2}+l^{2}\right] \delta _{l j}  \notag
\\
&&+\frac{t}{2}\left( 2l-1\right) \delta _{l j+1}+\left( \frac{t}{2}\right)
^{2}\delta _{l j+2}.  \label{s23}
\end{eqnarray}%
Finally, Eqs.(\ref{mome0}, \ref{mome}) are obtained substituting Eqs.(\ref%
{s13}, \ref{s23}) into Eqs.(\ref{unop}, \ref{dosp}).

\end{document}